\input{aipcheck}

\documentclass[
    ,final            
  ]{aipproc}

\layoutstyle{8x11single}
\newcommand{\eqb}{\begin{equation}}
\newcommand{\eqe}{\end{equation}}
\newcommand{\dmb}{\begin{displaymath}}
\newcommand{\dme}{\end{displaymath}}
\newcommand{\pd}{\partial}

\newcommand{\eab}{\begin{eqnarray}}
\newcommand{\eae}{\end{eqnarray}}

\newcommand{\e}{\mbox{e}}
\newcommand{\be}{\begin{equation}}
\newcommand{\ee}{\end{equation}}

\begin{document}

\title{Fundamental and effective Yang-Mills vertices}

\classification{}
\keywords{Calorons, effective gauge coupling, Planck's quantum of action}

\classification{11.10.Wx,11.15.Tk,11.55.Fv,02.60.Cb}

\author{Ralf Hofmann}{
  address={ITP, Universit\"at Heidelberg, Philosophenweg 16, 69120 Heidelberg, Germany}
}

\begin{abstract}

Calorons and plane waves within and in between them 
{\sl collectively} give rise to a thermal ground state. The latter provides 
a homgeneous energy density and a negative pressure, and it induces quasiparticle masses 
to part of the propagating spectrum of deconfining SU(2) Yang-Mills thermodynamics (dynamical gauge-symmetry breaking). 
In the present talk we discuss the role of a {\sl single} caloron in inducing 
effective local vertices, characterized by powers of $\hbar$, mediating the interaction of 
plane waves which propagate over large distances. The constraints on momentum 
transfers through effective 4-vertices are revisited.       
    
\end{abstract}

\maketitle


\section{Introduction}

Quantum Yang-Mills theory is a beautiful, successful, and useful concept which, 
in a largely perturbative interpretation, underlies the present Standard Model 
of particle physics based on the gauge group U(1)$\times$SU(2)$\times$SU(3). 
Its so-far analysed mathematical implications are impressive and include the construction of all 
(nonpropagating) selfdual solutions to the 4D Euclidean classical Yang-Mills equations of motion \cite{ADHM,Nahm1980,Nahm1981,Nahm1983},  
the topological characterization of knots \cite{Jones1985} in terms of Wilson-loop expectations 
in 3D Chern-Simons theory \cite{Witten1989} and of differentiable 4D manifolds \cite{Donaldson1990}, and investigations on the geometry and 
dynamics of minimal Yang-Mills solitons \cite{'tHooft1974,Polyakov1974,PrasadSommerfield1975,JuliaZee1975,Manton1982,AtiyahHitchin1988}. 
A description of nature in terms of gauge theory, which quantitatively accounts for the 
ground-state physics, dynamical gauge-symmetry breaking, the (residual) interaction of associated quasiparticles, and the transition to phases 
with effectively reduced gauge symmetries, requires the combined consideration of 
plane-wave fluctuations and nonpropagating, selfdual configurations. 
In a thermodynamic setting, we have generated results in this context 
\cite{HerbstHofmann2004-5,Hofmann2005,GiacosaHofmann2005,GiacosaHofmann2007,SchwarzHofmannGiacosa2007,LudescherHofmann2008,
Hofmann2009,Ludescheretal2008,FalquezHofmannBaumbach2010}, for a summary see \cite{HofmannBook}. A thermodynamical framework is natural 
since we observe an abundance and overlap of (unresolved) topological solitons of SU(2) Yang-Mills theory (magnetic monopoles) 
in the thermal ground-state \cite{Ludescheretal2008} which, in 
an out-of-thermal-equilibrium situation, leaves little hope for a deterministic description. 

In the present talk we 
elaborate on the correspondence between effective and fundamental field configurations in the deconfining phase 
of SU(2) Yang-Mills theory, for a discussion of the implications towards low-temperature 
photon physics see Markus Schwarz' talk \cite{MS}. In particular, we 
will justify the following picture of how the effective theory 
\cite{HerbstHofmann2004-5,Hofmann2005} describes the presence and interplay 
of particular selfdual field configurations, calorons of topological charge unity, and 
plane-wave fields: On one hand, those calorons, which in isolation possess vanishing 
energy-momentum, collectively form an inert and adjoint scalar field 
$\phi$. This is understood by computing a flat spatial average and a flat average over the caloron scale parameter $\rho$ of 
an adjointly transforming two-point correlator of the field strength to define the kernel ${\cal D}$ of a second-order differential 
operator which determines $\phi$'s equation of motion. The observation here is that ${\cal D}$ is dominated 
by calorons with $\rho\sim |\phi|$. Trivial-holonomy calorons \cite{HarringtonShepard1978} exhibit the potential to 
source and absorb collective plane-wave 
fluctuations with off-shellness greater than $|\phi|^2$ which, in turn, deform these calorons 
(trivial holonomy $\rightarrow$ small holonomy: short-lived magnetic dipoles \cite{Nahm1983,LeeLu1998-5,KraanVanBaal1998I-5,Diakonov2004}). 
Taking a coarse-grained version of this effect into account, an accurate estimate 
of the thermal ground-state emerges: Fundamental, collective plane-wave 
fluctuations contribute in terms of an effective pure-gauge configuration 
which induces pressure $P_{\tiny\mbox{gs}}$ and energy density $\rho_{\tiny\mbox{gs}}=-P_{\tiny\mbox{gs}}$ linear 
in temperature $T$. On the other hand, computing the gauge coupling $e$ in 
the effective theory and, by virtue of $\rho\sim |\phi|^{-1}$ for contributing calorons, using $e$ to determine its action $S_{\rho\sim |\phi|^{-1}}$, 
the results is $S_{\rho\sim |\phi|^{-1}}=\frac{8\pi^2}{e^2}=\hbar$ almost everywhere 
in the deconfining phase. This suggests 
that a local Yang-Mills vertex, which, after an appropriate rescaling of the propagating, effective gauge field, 
carries powers of $\hbar^{1/2}$ (3-vertex) or $\hbar$ (4-vertex), is {\sl induced} by the distortion of isolated, fundamental plane-wave 
propagations through a single caloron. {\sl Thus the calorons within the thermal ground-state lend themselves 
to induce vertices to the sector of propagating, isolated plane waves.} Compared to the contribution of freely propagating, 
effective plane waves this effect amounts to small radiative corrections (collective effect of 
screened magnetic monopoles generated by dissociation of large-holonomy calorons)
\cite{SchwarzHofmannGiacosa2007,LudescherHofmann2008} in thermodynamical quantities. Since only calorons with 
$\rho\sim |\phi|^{-1}$ are contributing to the thermal ground state momentum transfers through a vertex harder than $|\phi|^{2}$ 
are essentially suppressed. As a consequence, 
no ultraviolet divergences arise in integrating out isolated, fundamental plane-wave 
fluctuations harder than $|\phi|^2$ at resolution $|\phi|$. 
We stress that the perturbative renormalization programme is in agreement with this observation and 
that perturbative renormalizablity \cite{HooftVeltman,Zinn-Justin1972}, which states the form invariance of the Yang-Mills 
action under the process of perturbatively integrating out plane-wave fluctuations that are ultraviolet relative to 
a given renormalization point $\mu$, is an important guide in the construction 
of the effective action in the deconfining phase.

\section{Effective Yang-Mills action and action of a contributing caloron}

The spatial coarse-graining over noninteracting trivial-holonomy calorons \cite{HarringtonShepard1978} is 
performed by determining, in the singular gauge used to construct this caloron in terms of the prepotential of a 
charge-one instanton \cite{'tHooft1976U,JackiwRebbi1976}, the second-order linear operator 
$D\equiv\partial^2_\tau+\left(\frac{2\pi}{\beta}\right)^2$ which annihilates the 
field $\phi$ \cite{HerbstHofmann2004-5,Hofmann2005,GiacosaHofmann2007}: $D\phi=0$. 
To learn about $D$ in terms of its kernel ${\cal K}$, a flat spatial and flat $\rho$ average over a unique, adjointly transforming, scalar two-point correlator 
of the Yang-Mills field strength needs to be performed 
\cite{HerbstHofmann2004-5}. In this process we observe that 
(i) only magnetic-magnetic correlations contribute and 
that (ii) the $\rho$ integration and thus ${\cal K}$ is saturated essentially by contributions around 
the upper limit $\rho_c=|\phi|^{-1}$ since the integral 
scales as $\rho_c^3$ \cite{HerbstHofmann2004-5,HofmannKaviani2012}. 
Demanding consistency of $D\phi=0$ 
with $\phi$'s first-order Bogmoln'yi equation determines the potential $V(\phi)$ 
in the effective action. $V(\phi)$ is unique modulo a {\sl multiplicative} constant: $V(\phi)=\frac{\Lambda^6}{\phi^2}$. 
In contrast to perturbation theory, the Yang-Mills scale $\Lambda$ enters here on the level of nonpropagating field configurations 
as a constant of integration for the following first-order consistency equation \cite{GiacosaHofmann2007}
\eqb
\label{eomPot}
\frac{\partial V(|\phi|^2)}{\partial
  |\phi|^2}=-\frac{V(|\phi|^2)}{|\phi|^2}\,,
\eqe   
where $|\phi|$ denotes the space-time independent and 
gauge invariant modulus of $\phi$. Invoking perturbative renormalizability \cite{HooftVeltman,Zinn-Justin1972} 
and the fact that the (nonpropagating) field $\phi$ is inert, one arrives at the unique effective action density ${\cal L}_{\mbox{\tiny eff}}$, 
for details see \cite{HofmannBook}: 
\begin{equation}
\label{fullactden}
{\cal L}_{\mbox{\tiny eff}}[a_\mu]=\mbox{tr}\,\left(\frac12\,
  G_{\mu\nu}G_{\mu\nu}+(D_\mu\phi)^2+\frac{\Lambda^6}{\phi^2}\right)\,,
\end{equation} 
where $G_{\mu\nu}=\partial_\mu a_\nu-\partial_\nu
a_\mu-ie[a_\mu,a_\nu]\equiv G^a_{\mu\nu}\,t_a$ denotes the field 
strength for the effective gauge field $a_\mu$, $D_\mu\phi=\partial_\mu\phi-ie[a_\mu,\phi]$, and $e$ is the effective
gauge coupling to be determined, see below. ${\cal L}_{\mbox{\tiny eff}}$ in Eq.\,(\ref{fullactden}) yields a highly accurate tree-level 
ground-state estimate and, as easily deduced in unitary gauge $\phi=2|\phi|\,t_3$, tree-level 
mass $m=2\,e|\phi|=2\,e\sqrt{\frac{\Lambda^3}{2\pi T}}$ for propagating 
gauge modes $a_\mu^{1,2}$ thus representing thermal quasiparticles. The emergence of mass in the propagation of effective plane waves 
relates to the {\sl collective} effect of calorons residing in the thermal ground state, below we are 
interested in the induction of effective and local vertices by {\sl single} calorons.   

Demanding Legendre transformations between thermodynamical quantities (pressure, energy density etc.), 
which are a consequence of the temperature independence of the parameters of the fundamental action (infinite resolution), 
on the level of free thermal quasiparticles \footnote{In the physical unitary-Coulomb gauge $\phi=2|\phi|\,t_3\,,\ \pd_i a_i^3=0$ the 
off-shellness of quantum fluctuations in the field $a_\mu$ is constrained by the maximal resolution 
$|\phi|$. Their contribution turns out to be negligible compared to 
that of thermal fluctuations \cite{Hofmann2005}.}, one obtains the following evolution equation \cite{Hofmann2005}
 \begin{equation}
\label{evoleqsu2}
\partial_a\lambda=-\frac{24\lambda^4
  a}{(2\pi)^6}\frac{D(2a)}{1+\frac{24\lambda^3a^2}{(2\pi)^6}D(2a)}\,,
\end{equation}
where $\lambda\equiv\frac{2\pi T}{\Lambda}$, $a\equiv\frac{m}{2T}=2\pi e\lambda^{-3/2}$, 
and $D(y)\equiv \int_0^\infty dx\,\frac{x^2}{\sqrt{x^2+y^2}}\,\frac{1}{\e^{\sqrt{x^2+y^2}}-1}$. 
For a high-temperature boundary 
condition $a(\lambda_i)=a_i\ll 1$, the solution to Eq.\,(\ref{evoleqsu2}) is approximately given as \cite{HofmannBook}
\begin{equation}
\label{solasu2} 
a(\lambda)=4\sqrt{2}\pi^2\lambda^{-3/2}\left(1-\frac{\lambda}{\lambda_i}\left[1-\frac{a^2_i\lambda_i^3}{32\pi^4}\right]\right)^{1/2}\,.
\end{equation}
Thus for sufficiently low $\lambda_i$ function $a(\lambda)$ runs into the
attractor $a(\lambda)=4\sqrt{2}\pi^2\lambda^{-3/2}$. Using $a\equiv\frac{m}{2T}=2\pi
e\lambda^{-3/2}$, this attractor corresponds to the plateau $e\equiv\sqrt{8}\pi$. 
Lowering $\lambda$ beyond a point, where $a\ll 1$ holds, the true solution to 
Eq.\,(\ref{evoleqsu2}) runs into a thin pole at $\lambda_c$ of the form $e(\lambda)=-4.59\,\log(\lambda-\lambda_c)+18.42$ with 
$\lambda_c=13.87$ \cite{HofmannBook}. The logarithmic pole of $e$ at $\lambda_c$ initiates a transition to a phase 
where thermal quasiparticles decouple (relevant for a higgsfree breaking of electroweak gauge symmetry \cite{HofmannBook}), 
the ground state starts to rearrange into a {\sl condensate} of magnetic monopoles, and 
the formerly massless gauge mode picks up mass by the dual Meissner effect \cite{Hofmann2005}. We conclude that {\sl almost everywhere} 
in the deconfining phase the effective coupling assumes the constant value 
$e\equiv\sqrt{8}\pi$. 

Let us now present the usual argument for counting the powers in $\hbar$ associated with vertices 
in the effective theory. We work in 
units where $k_B=c=\epsilon_0=\mu_0=1$ but $\hbar$ is re-instated as an action. (So far we have 
worked in supernatural units $\hbar=c=k_B=1$). 
The (dimensionless) exponential 
\eqb
\label{expweight}
-\frac{\int_0^{\beta} d\tau d^3x\,{\cal L}^\prime_{\mbox{\tiny eff}}[a_\mu]}{\hbar}\,, 
\eqe
in the weight belonging to fluctuating fields in the partition function can, in unitary gauge, be re-cast as 
\eqb
\label{expweightdiml}
-\int_0^{\beta} d\tau d^3x\,\mbox{tr}\,
\Big(\frac12(\partial_\mu \tilde{a}_\nu-\partial_\nu
\tilde{a}_\mu-ie\sqrt{\hbar}[\tilde{a}_\mu,\tilde{a}_\nu])^2-e^2\hbar[\tilde{a}_\mu,\tilde{\phi}]^2\Big)\,,
\eqe
where $\tilde{a}_\mu\equiv a_\mu/\sqrt{\hbar}$ and $\tilde{\phi}\equiv \phi/\sqrt{\hbar}$ 
are assumed indepent of $\hbar$ \cite{Brodsky2011}, see also \cite{DonoghuePapers}. Notice that because of the terms  
$\propto \hbar^0$ in (\ref{expweightdiml}) the unit of $\tilde{a}_\mu$ is 
length$^{-1}$. Thus the coupling $e$ must have the unit of $1/\sqrt{\hbar}$ 
(and the unit of $\tilde{\phi}$ is also length$^{-1}$). As a result, we have $e\equiv\frac{\sqrt{8}\pi}{\sqrt{\hbar}}$ 
almost everywhere in the deconfining phase. 
Because mainly (just-not-resolved) calorons with $\rho\sim |\phi|^{-1}$ contribute to the effective theory, 
the action $S_{\rho\sim |\phi|^{-1}}$ of each such calorons reads
\eqb
\label{actioncalantcal}
S_{\rho\sim |\phi|^{-1}}=\frac{8\pi^2}{e^2}=\hbar\,.
\eqe
Equation (\ref{actioncalantcal}) has implications:\\ 
\noindent First, it states that Planck's quantum of action 
$\hbar$, whose introduction in (\ref{expweight}) ultimately has its origin in the laws of 
Quantum Mechanics and which accounts for quantum 
corrections (number of loops) in the effective theory, coincides with 
the Euclidean action of a just-not-resolved selfdual (classical) field configuration. That is, the reason why 
$\hbar$ really is a constant can be traced to the universal constancy of $e$ (almost no dependence on $T$ and 
on the specific Yang-Mills scale $\Lambda$ of a given 
SU(2) gauge theory) and the 
fact that there is one unit of conserved topological charge to a caloron. 
Moreover, rather than multiplying first and second 
powers of $\hbar^{1/2}$ onto the powers $e$ and $e^2$ 
of an a priori unknown coupling constant $e$ appearing in the three- and four-vertices of a Yang-Mills 
theory we would infer that these vertices are {\sl induced} by just-not-resolved individual 
calorons.\\ 
\noindent Second, for the fine-structure constant $\alpha$ of 
Quantum Electrodynamics (QED) to be dimensionless,  
\eqb
\label{FSC}
\alpha=N^{-1}\frac{g^2}{4\pi\hbar}\,,
\eqe
the coupling $g$ in Eq.\,(\ref{FSC}) must have the unit of $\sqrt{\hbar}$. ($N^{-1}$ denotes a numerical 
factor that should relate to the mixing of the massless modes belonging to several SU(2) 
groups, see \cite{GiacosaHofmann2005}.) This happens if $g$ is taken to be the electric-magnetically dual coupling 
to $e$: 
\eqb
\label{elch}
g=\frac{4\pi}{e}\propto \sqrt{\hbar}\,.
\eqe
That is, a {\sl magnetic} monopole liberated by the dissociation of a large-holonomy SU(2) caloron and other incarnations of this 
magnetic charge in the preconfining and confining phase\footnote{For example, the isolated charge situated in the center of the flux 
eddy associated with the locus of selfintersection of a center-vortex loop in the confining phase 
\cite{HofmannBook,MoosmannHofmann2008}.} are interpreted as {\sl electric} charges in the real world. Third, 
by virtue of $e\equiv\frac{\sqrt{8}\pi}{\sqrt{\hbar}}$ the factors $e\sqrt{\hbar}$ and $e^2\hbar$ in (\ref{expweightdiml}) 
do not depend on $\hbar$. Thus, by the above assumption that the fields $\tilde{a}_\mu,\tilde{\phi}$ are independent of $\hbar$, 
the weight (\ref{expweight}) is also independent of 
$\hbar$, and thus the effective loop expansion is {\sl not} an expansion in powers of $\hbar$. This is in contrast to renormalized perturbation 
theory where the value of the coupling is assumed to be independent of $\hbar$.

\section{Momentum transfers in an effective 4-vertex}

According to the Feynman rules associated with the effective theory (\ref{fullactden}) the 4-vertex is local and reads
\begin{eqnarray}
\Gamma^{\mu\nu\rho\delta}_{[4]abcd} & = & -ie^2(2\pi)^4\delta(p+q+s+r)
[\varepsilon_{fab}\varepsilon_{fcd}(g^{\mu\rho}g^{\nu\sigma}-g^{\mu\sigma}g^{\nu\rho})+ \nonumber  \\
&& \varepsilon_{fac}\varepsilon_{fdb}(g^{\mu\sigma}g^{\rho\nu}-g^{\mu\nu}g^{\rho\sigma})+ 
\varepsilon_{fad}\varepsilon_{fbc}(g^{\mu\nu}g^{\sigma\rho}-g^{\mu\rho}g^{\sigma\nu})]\,,\nonumber\\ 
\end {eqnarray}
where $\varepsilon_{abc}$ are the structure constants of SU(2). We work in physical unitary-Coulomb gauge and in a real-time re-formulation 
of the thermodynamics \cite{HofmannBook}. A 2$\rightarrow$ 2 effective amplitude, where $\Gamma^{\mu\nu\rho\delta}_{[4]abcd}$
links incoming four-momenta $p_1$ and $p_2$ with outgoing four-momenta $p_3$ and $p_4=p_1+p_2-p_3$, 
contains all fundamental scattering channels $s=(p_1+p_2)^2$, $t=(p_3-p_1)^2$, and $u=(p_2-p_3)^2$. This is expressed by the fact that 
the according effective tree-diagram subject to the local vertex $\Gamma^{\mu\nu\rho\delta}_{[4]abcd}$ 
is invariant under permutations of the external legs. For the inducing caloron to 
remain unresolved in the effective theory we need to demand that $s,t,u\le |\phi|^2$. 
Notice that for (naive) on-shellness, either 
$p_1^2=m^2$, $p_2^2=m^2$, $p^2_3=m^2$, $p^2_4=m^2$ or $p_1^2=m^2$, $p_2^2=m^2$, $p^2_3=0$, $p^2_4=0$, the conditions 
$s\le |\phi|^2$, $t\le |\phi|^2$, and $u\le |\phi|^2$ cannot be satisfied simultaneously\footnote{The author would like to thank 
Niko Krasowski for pointing this out.}.  A channel $s$ or $t$ or $u$ is 
called trivial if $s=0$ or $t=0$ or $u=0$, respectively. Let $N=1\cdots 3$ be the number of nontrivial channels. 
Writing $\Gamma^{\mu\nu\rho\delta}_{[4]abcd}=\frac{1}{N}(\left.\Gamma^{\mu\nu\rho\delta}_{[4]abcd}\right|_s+\cdots+
\left.\Gamma^{\mu\nu\rho\delta}_{[4]abcd}\right|_u)$,  
we associate a nontrivial channel with each summand and assign the according constraint. For example, 
in the 4-vertex diagram 
for the one-loop polarization tensor of the massless mode ($p$ the external and $k$ the loop momentum) 
the $t$ channel is trivial and thus $N=2$. Also, if  
$s\le |\phi|^2$ for a given combination $p$ and $k$ then $u>|\phi|^2$ thus excluding the $u$-channel. 
Alternatively, $u\le|\phi|^2$ for $p$ and $-k$ and 
then $s>|\phi|^2$ thus excluding the $s$-channel. 
Since the integrand, associated with the $k$-loop, is invariant under $k\to -k$ one effectively 
has  
\eqb
\label{effvertexpol}
\Gamma^{\mu\nu\rho\delta}_{[4]abcd}=\frac{1}{2}(\left.\Gamma^{\mu\nu\rho\delta}_{[4]abcd}\right|_s+
\left.\Gamma^{\mu\nu\rho\delta}_{[4]abcd}\right|_u)=\left.\Gamma^{\mu\nu\rho\delta}_{[4]abcd}\right|_s\,.
\eqe
Eq.\,(\ref{effvertexpol}) holds equally well for the computation of two-loop corrections to the pressure 
that invoke one 4-vertex. (Here the two cases $p,k$ and $-p,-k$, where the $s$-channel constraint is 
satisfied (violated), are complemented by the two cases $-p,k$ and $p,-k$, where the $u$-channel constraint is 
violated (satisfied) while the loop-integrands are invariant under $p\to -p$ and  $k\to -k$.)

\section{Summary} 

In this talk we have argued on thermodynamic grounds for a representation of Planck's quantum 
of action $\hbar$ in terms of the Euclidean action of an isolated, just not resolved 
SU(2) Yang-Mills caloron of topological charge unity. Such a caloron acts as an 
inducer of a local 3-vertex or 4-vertex in the effective theory. We have also re-discussed how the 
momentum transfer in a $2\to 2$ amplitude, conveyed by such a caloron, is to be treated 
in the effective theory where the accoding Feynman rule is blind towards changes of the 
scattering channel.

\section*{Acknowledgments}
The author would like to thank Markus Schwarz and Niko Krasowkski for discussions.


\end{document}